\documentclass{moriond}
\usepackage[english]{babel}
\usepackage[utf8]{inputenc}
\usepackage[T1]{fontenc}
\usepackage{etoolbox}
\usepackage{physics}
\usepackage{verbatim}
\usepackage{mathtools}
\usepackage{amsmath}
\usepackage{amssymb}

\def\Journal#1#2#3#4{{#1} {\bf #2}, #3 (#4)}

\def\PLB{{\em Phys. Lett.}  B}

\def\PRD{{\em Phys. Rev.} D}

\def\LRR{{\em Living Rev. Rel.}}
\def\CQG{{\em Class. Quant. Grav.}}

\def\be{\begin{equation}}
\def\ee{\end{equation}}
\def\bea{\begin{eqnarray}}
\def\eea{\end{eqnarray}}

\newcommand{\Photo}{}

\begin{document}
\vspace*{4cm}
\title{PQM and the GUP: Implications of Lattice Dynamics and Minimal\\Uncertainties in Quantum Mechanics and Cosmology}

\author{G. Barca, E. Giovannetti}
\address{Department of Physics, “La Sapienza” University of Rome, P.le Aldo Moro 5 (00185) Rome, Italy.}

\author{G. Montani}
\address{ENEA, Fusion and Nuclear Safety Department, Via E. Fermi, 45 (00044) Frascati (RM), Italy;\\Department of Physics, “La Sapienza” University of Rome, P.le Aldo Moro 5 (00185) Rome, Italy.}

\maketitle

\abstracts{
We compare two alternative representations of quantum mechanics: Polymer Quantum Mechanics (PQM), which presents features similar to Loop Quantum Gravity and Loop Quantum Cosmology, and the Generalized Uncertainty Principle (GUP) representation, that through a modified Algebra yields an alternative uncertainty principle similar to those derived in String Theories and Brane Cosmology. These formalisms can be recast to apparently look similar, but while the GUP yields an absolute minimal uncertainty on position, PQM implements some kind of ultraviolet cut-off through a lattice and does not have a minimal uncertainty. Then we implement them on the anisotropic Bianchi I model in Misner-like variables on a semiclassical level: PQM always implies a removal of the singularities, while the GUP fails to do so, highlighting once again how the two representations are fundamentally incompatible.}

\section{Introduction}
The two most developed Quantum Gravity theories are Loop Quantum Gravity (LQG) \cite{LQG} and String Theories (STs) \cite{Stringhe}. The cosmological implementation of LQG, known as Loop Quantum Cosmology (LQC) \cite{LQC1,LQC2}, is well reproduced by the minisuperspace implementation of Polymer Quantum Mechanics (PQM) \cite{PQM}. Similarly, the low energy phenomenology of STs can be reproduced by the Generalized Uncertainty Principle representation (GUP) \cite{Maggiore1,Maggiore2,Kempf} which, when applied to the minisuperspace, yields a dynamics similar to Brane Cosmology (BC) \cite{BraneCosmology}.

Here we present a comparison between the implementation of the two phenomenological approaches of PQM and GUP on the Bianchi I model. The leading idea is that both approaches can be stated in terms of modified Heisenberg algebras, with the important difference of a sign.

We use the units $\hslash=c=8\pi G=1$. For more details on this work, see the original paper \cite{nostro}.

\section{Alternative Representations of Quantum Mechanics}
\subsection{Polymer Quantum Mechanics}
PQM reproduces effects similar to LQG through a simpler, independent framework \cite{PQM}; by assigning one variable a discrete character, the conjugate momentum cannot be promoted to a well-defined quantum operator and must be regularized through the introduction of a lattice on position with constant spacing $\mu_0$. Then the regularized momentum is a multiplicative operator of the form
\begin{equation}
    \hat{p}\,\psi(p)=\frac{\sin(\mu_0p)}{\mu_0}\,\psi(p).
    \label{ppoly}
\end{equation}

\subsection{The Generalized Uncertainty Principle Representation}
The GUP is an extension of the standard Heisenberg Uncertainty Principle inspired by STs. Introducing modified commutation relations \cite{Maggiore1,Maggiore2}, it is possible to reproduce the modified Uncertainty Principle that appears in STs, which implies an absolute minimal uncertainty on position:
\begin{equation}
    \comm{q}{p}=i\,(1+Bp^2)\qq{with}B>0,\quad\Delta q\Delta p=\frac{1}{2}\,\Big(1+B(\Delta p)^2\Big),\quad\Delta q_0=\sqrt{B\,}\,.
    \label{kempfcomm}
\end{equation}

\subsection{PQM as a Modified Algebra}
It is possible to implement PQM effects also through the modified algebra \cite{Battisti} $\comm{q}{p}=i\,\sqrt{1-\mu_0^2p^2\,}\,$; starting from this, we performed two different analyses. Firstly we Taylor-expanded the square root and truncated (“t”) at second order to obtain a commutator similar to the GUP one \eqref{kempfcomm}; in the second case, we used the exact (“e”) polymer representation \eqref{ppoly} inside the square root.
\begin{equation}
    \comm{q}{p}_\text{t}=i\,\left(1-\frac{\mu_0^2}{2}p^2\right),\quad(\Delta q)^2_\text{t}\geq\frac{\mu_0^4}{4}\ev{p}^2-\frac{\mu_0^2}{2};\quad\comm{q}{p}_\text{e}=i\,\cos(\mu_0p),\quad(\Delta q)_\text{e}\geq\frac{e^{-\frac{\mu_0^2(\Delta p)^2}{4}}}{4\,(\Delta p)^2\sqrt{\pi\,}}.
    \label{commpup}
\end{equation}
In both these Polymer Uncertainty Principles (PUPs) there is no minimal uncertainty on position at zero order, and thus we can say that PQM and the GUP are intrinsically different.

\subsection{A Different GUP formulation for Brane Cosmology}
Now, from the original GUP commutator \eqref{kempfcomm} we can construct a new modified commutator; the corresponding uncertainty principle still implies (at zero order) an absolute minimal uncertainty:
\begin{equation}
    \comm{q}{p}_\text{b}=i\,\cosh(\sqrt{2B\,}\,p),\quad\Delta q\Delta p\geq\frac{e^{\frac{B(\Delta p)^2}{2}}}{4\,\Delta p\,\sqrt{\pi\,}},\quad\Delta q_0=\frac{eB}{8\sqrt{\pi\,}}.
\end{equation}
This formalism has been named brane-GUP (“b”) since it reproduces Brane Cosmology (BC).

\subsection{Formalisms for the Semiclassical Cosmological Implementation}
PQM can be easily implemented on any Hamiltonian system through a modified Hamiltonian function where the momentum has been replaced by the sine function of equation \eqref{ppoly}.

The GUP and PUPs representations can be implemented using modified Poisson brackets that are the classical limits of the modified commutators \eqref{kempfcomm} and \eqref{commpup} respectively.

The GUPb formulation will be implemented in two ways: by using modified Poisson brackets $\pb{q}{p}=\sqrt{1-2Bp^2\,}$; or by constructing a modified Hamiltonian where, similarly to PQM, the momentum has been replaced by a hyperbolic sine function.

\section{Bianchi I Semiclassical Dynamics}
Now we implement all the formulations introduced above to the Bianchi I anisotropic model.

We use a Hamiltonian formalism and a variation of the Misner variables: the isotropic volume $v$ and the anisotropies $\beta_\pm$; we also introduce a free massless scalar field $\phi$. From the equations of motion we derive the Friedmann equations at first in synchronous time (N$=1$); then we perform a change of timegauge setting $\dot{\phi}=1$ and solve the dynamics by finding the evolution $v(\phi)$.

\subsection{Classical Dynamics}
First we show the classical dynamics. The Hamiltonian constraint and Friedmann equation are
\begin{equation}
    \mathcal{C}_\text{class}=\frac{-9P_v^2v^2+P_+^2+P_-^2}{\xi\,v}+\frac{\xi}{12}\,\rho_\phi v=0;\quad H^2=\left(\frac{\dot{v}}{3v}\right)^2=\frac{\rho_\phi+\rho_\text{a}}{3};
\end{equation}
we see that the anisotropies contribute to the total energy density through $\rho_\text{a}\propto\frac{P_+^2+P_-^2}{v^2}$.

By changing the timegauge and rewriting the Friedmann equation, its r.h.s. results to be a constant; it yields two singular solutions of the form $v(\phi)=v_0\,e^{\pm v_1\phi}$ : the plus sign is a model expanding from a Big Bang, while the minus sign is a Universe contracting towards a Big Crunch.

\subsection{Modified Bianchi I Cosmology}
Now we present the semiclassical implementation of the alternative quantum mechanical representations introduced above. Note that we modify only the volume sector, leaving the anisotropies and the scalar field untouched. For each formalism, we report the Hamiltonian constraint and, if necessary, the modified Poisson brackets that we used to derive the dynamics.
\begin{equation}
    \text{PQM:}\quad\mathcal{C}_\text{PQM}=\frac{1}{\xi}\left(-9v\,\frac{\sin[2](\mu_0P_v)}{\mu_0^2}+\frac{P_+^2+P_-^2}{v}\right)+\frac{\xi}{12}\,\rho_\phi v=0;
\end{equation}
\begin{equation}
    \text{GUP:}\quad\mathcal{C}_\text{GUP}=\mathcal{C}_\text{class},\quad\pb{v}{P_v}_\text{GUP}=1+BP_v^2;
\end{equation}
\begin{equation}
    \text{PUP:}\quad\mathcal{C}_\text{PUP}=\mathcal{C}_\text{class},\quad\pb{v}{P_v}_\text{t}=\left(1-\frac{\mu_0^2}{2}P_v^2\right),\quad\pb{v}{P_v}_\text{e}=\sqrt{1-\mu_0^2P_v^2\,};
\end{equation}
\begin{equation}
    \text{GUPb1:}\quad\mathcal{C}_{\text{b1}}=\frac{1}{\xi}\left(-9v\,\frac{\sinh[2](\sqrt{2B\,}\,P_v)}{2B}+\frac{P_+^2+P_-^2}{v}\right)+\frac{\xi}{12}\,\rho_\phi v=0;
\end{equation}
\begin{equation}
    \text{GUPb2:}\quad\quad\quad\mathcal{C}_{\text{b2}}=\mathcal{C}_\text{class},\quad\pb{v}{P_v}_{\text{b2}}=\sqrt{1+2BP_v^2\,}.
\end{equation}

In Table \ref{semiclcosmology} we report the modified Friedmann equations and the corresponding solutions.
\begin{table}[h!]
    \centering
    \def\arraystretch{1.3}
    \hspace*{-0.8cm}\begin{tabular}{c|c|c|}
    &Mod. Hamiltonian or Poisson&Mod. Poisson brackets\\
    &brackets with square root&with quadratic term\\
    \hline
    PQM \& PUPs&$H^2=\frac{\rho_\phi+\rho_\text{a}}{3}\left(1-\frac{\rho_\phi+\rho_\text{a}}{\rho_\mu}\right)$&$H^2=\frac{\rho_\phi+\rho_\text{a}}{3}\left(1-\frac{\rho_\phi+\rho_\text{a}}{2\rho_\mu}\right)^2$\\
    (non-singular)&$v(\phi)=R_\mu\cosh(v_1\phi)$&$v(\phi)=R_\mu\sqrt{\frac{e^{\pm2v_1\phi}+1\,}{2}\,}$\\
    \hline
    GUP \& GUPb&$H^2=\frac{\rho_\phi+\rho_\text{a}}{3}\left(1+\frac{\rho_\phi+\rho_\text{a}}{\rho_B}\right)$&$H^2=\frac{\rho_\phi+\rho_\text{a}}{3}\left(1+\frac{\rho_\phi+\rho_\text{a}}{2\rho_B}\right)^2$\\
    (singular)&$v(\phi)=\pm R_B\sinh(v_1\phi)$&$v(\phi)=R_B\sqrt{\frac{e^{\pm2v_1\phi}-1\,}{2}\,}$\\
    \hline
    \end{tabular}
    \caption{Summary of the modified Friedmann equations and of the resulting modified evolutions $v(\phi)$.}
    \label{semiclcosmology}
\end{table}

First, all of them yield a correction factor proportional to $(1\pm\frac{\rho}{\rho_i})$ where $\rho_i$ are constant regularizing densities: $\rho_\mu=\frac{108}{\mu_0^2\xi^2}=\text{const.}$, $\rho_B=\frac{54}{B\xi^2}=\text{const.}$ This happens only using the volume $v$ as a configurational variable; in any other variable they would not be constant \cite{Mantero}.

\begin{figure}
    \centering
    \includegraphics[width=0.32\textwidth]{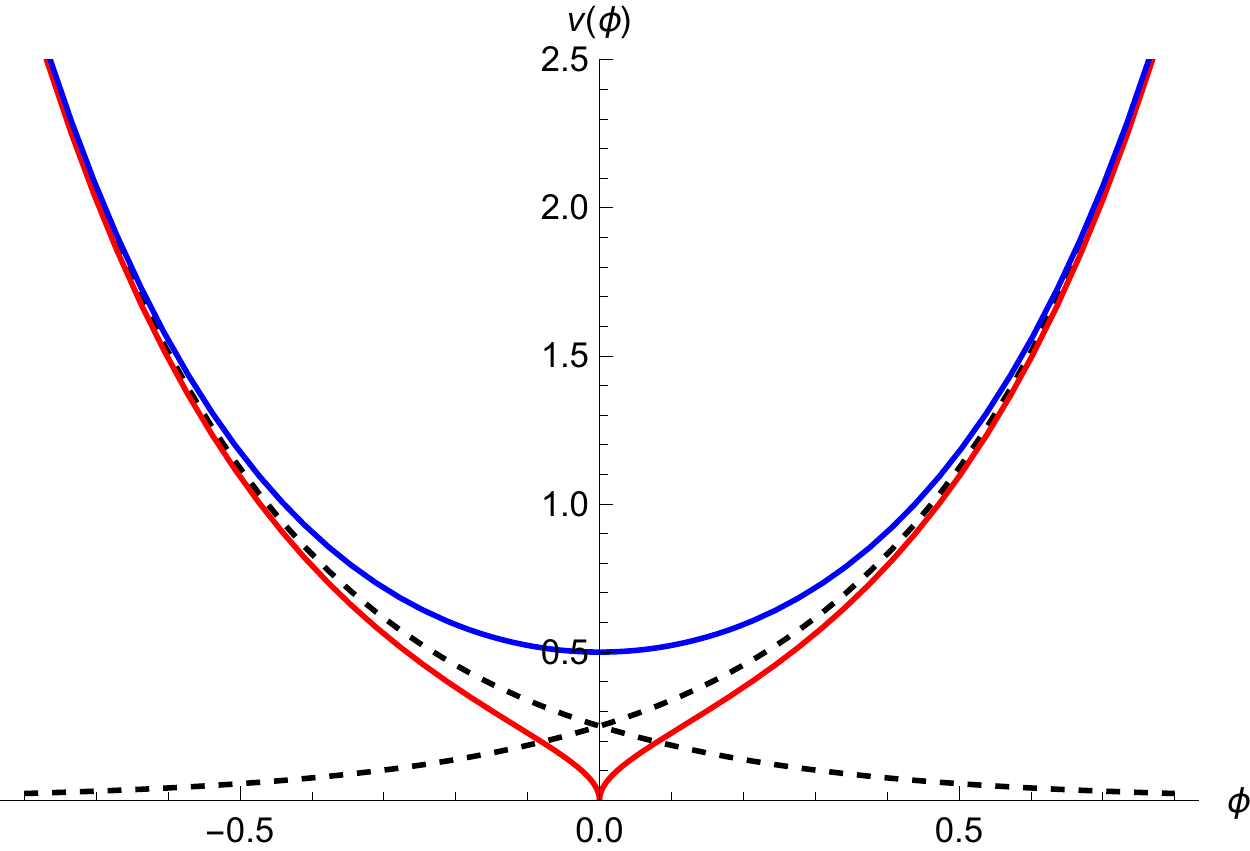}
    \,
    \includegraphics[width=0.32\textwidth]{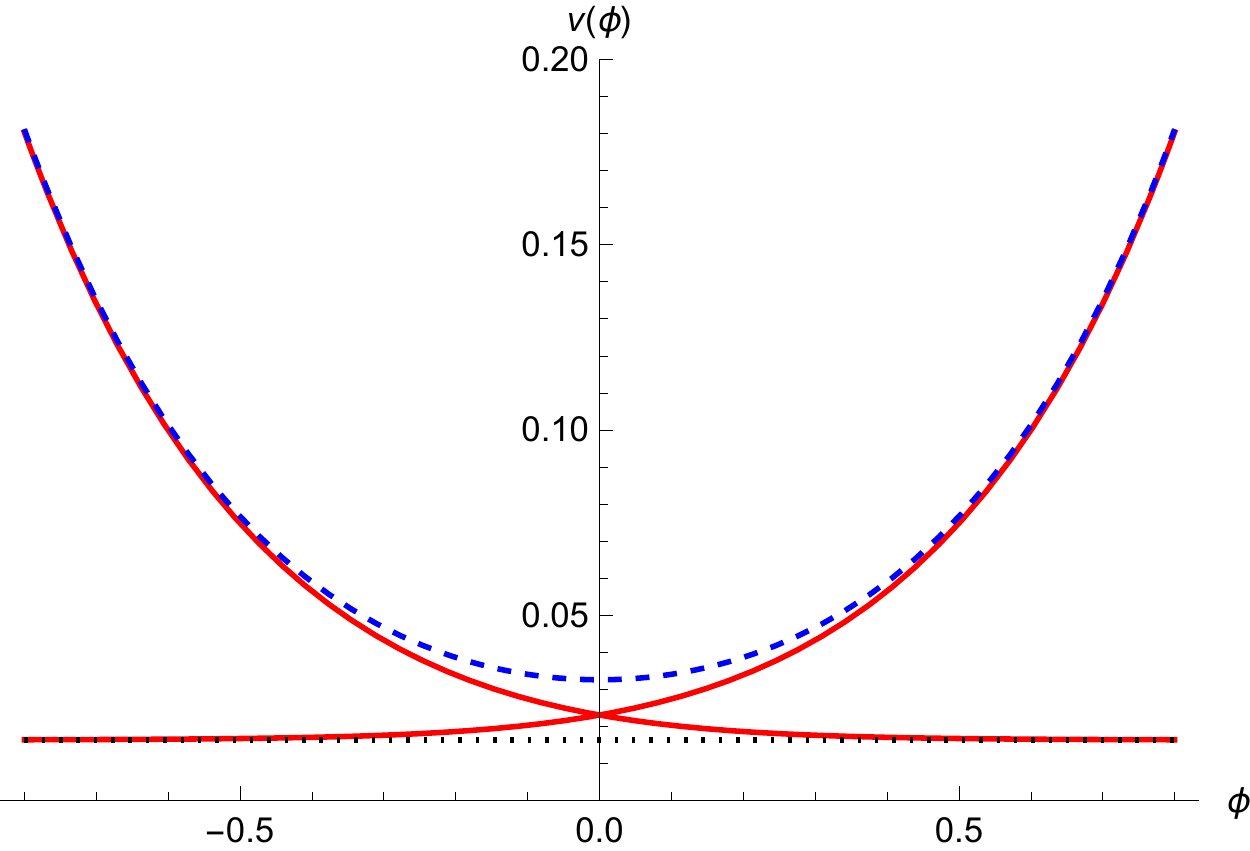}
    \,
    \includegraphics[width=0.32\textwidth]{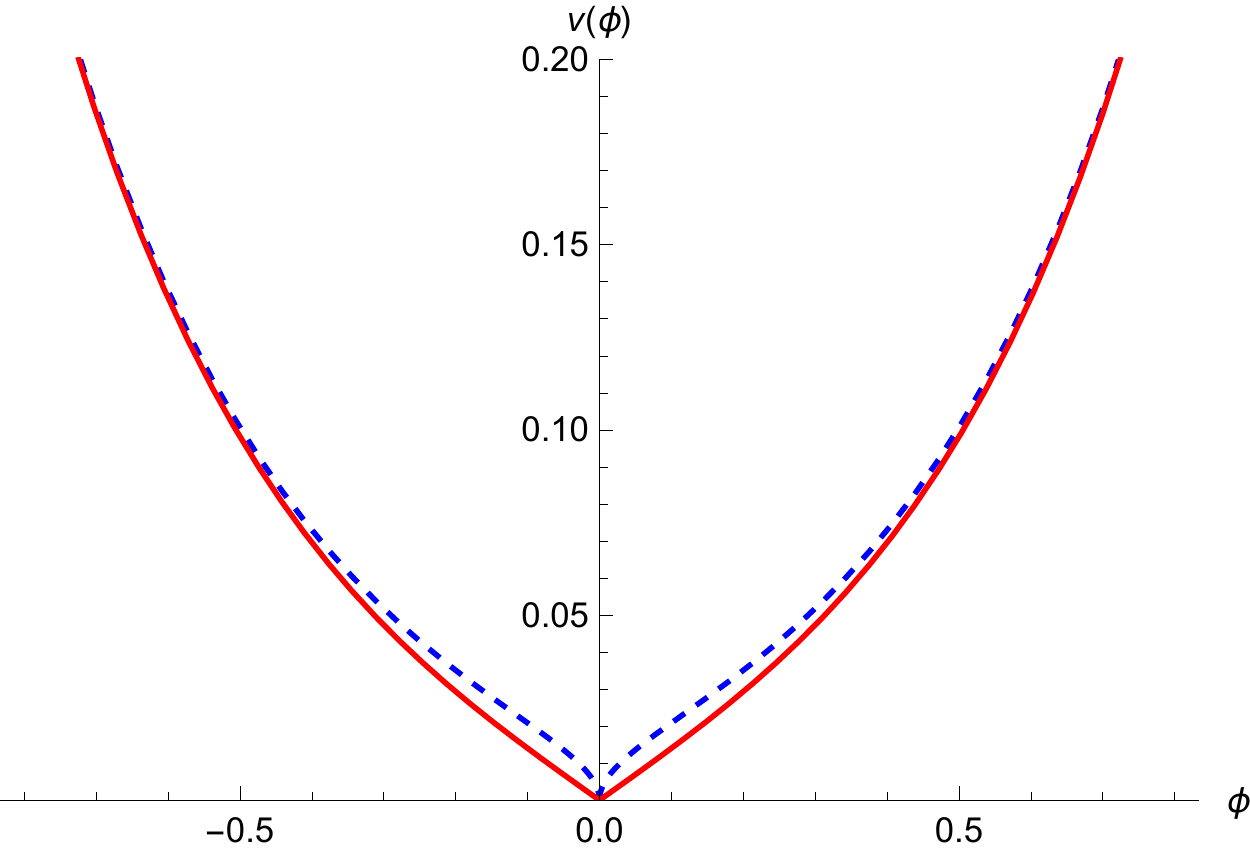}
    \caption{$v$ as function of time $\phi$. Left panel: comparison of PQM (blue) and GUP (red) dynamics with the classical solutions (black). Central panel: comparison of the PUPt asymptotic solutions (red) with the PQM Bouncing dynamics (blue). Right panel: similarity between the GUPb (red) and the original GUP (blue) solutions.}
    \label{moddynamics}
\end{figure}

Secondly, lattice corrections (i.e. PQM and the PUPs) always have a minus sign; this introduces a critical point in the dynamics and these formulations avoid the singularities, either through a Big Bounce (similarly to LQC) or through asymptotes, as shown in the left and central panels of Figure \ref{moddynamics}. The correction factors in all GUP formulations instead contain a plus sign, and thus their solutions are always singular, as shown in the left and right panels of Figure \ref{moddynamics}.

Another feature is that when quantum corrections are implemented through modified Poisson brackets with a quadratic term, the modified Friedmann equation has a $\rho^3$ leading term; if instead we use Poisson brackets with a square root or a Hamiltonian constraint modified with a sine function (trigonometric for PQM, hyperbolic for the GUPb), the leading term is $\pm\rho^2$.

Finally, the GUPb Friedmann equation is the same of BC \cite{BraneCosmology} (except for the curvature term).

We also implemented PQM and the GUP on a quantum level in their original formulations. We studied the evolution of wavepackets and found that the expectation values of the volume follow the corresponding semiclassical solutions. For more information, see the original paper \cite{nostro}.

\section{Summary and Concluding Remarks}
The two alternative representations of Quantum Mechanics presented here implement a fundamental minimal length in very different ways: PQM through a lattice and the GUP through an absolute minimal uncertainty. They can be reformulated to perform a comparison, but PQM never implies a minimal uncertainty, which is instead the main feature of the GUP.

The apparent similarity is present also in the minisuperspace implementation, but the evolution is very different: PQM always avoids the singularities, while the GUP solutions cannot.

To conclude, the GUP reproduces BC only if formulated with a square root or a hyperbolic sine. Therefore these formulations may be more faithful to the parent theories of ST and BC \cite{Maggiore2021}.


\section*{References}

\begin{thebibliography}{11}
\bibitem{LQG}C. Rovelli, {\em Quantum Gravity}
(Cambridge University Press, Cambridge, 2004).
\bibitem{Stringhe}M. Green, J. Schwarz and E. Witten, {\em Superstring Theory} (Cambridge University Press, Cambridge, 1987)
\bibitem{LQC1}M. Bojowald, \Journal{\LRR}{8}{11}{2005}.
\bibitem{LQC2}A. Ashtekar and P. Singh, \Journal{\CQG}{28}{213001}{2011}.
\bibitem{PQM}A. Corichi, T. Vukašinac and J.A. Zapata, \Journal{\PRD}{76}{044016}{2007}.
\bibitem{Maggiore1}M. Maggiore, \Journal{\PRD}{304}{65}{1993}.
\bibitem{Maggiore2}M. Maggiore, \Journal{\PLB}{49}{5182}{1994}.
\bibitem{Kempf}A. Kempf, G. Mangano and R.B. Mann, \Journal{\PRD}{52}{1108}{1995}.
\bibitem{BraneCosmology} R. Maartens and K. Koyama, \Journal{\LRR}{13}{5}{2010}.
\bibitem{nostro}G. Barca, E. Giovannetti and G. Montani, \Journal{preprint}{\hspace{-0.14cm}}{}{2021}
\bibitem{Battisti}M.V. Battisti, \Journal{\PRD}{79}{083506}{2009}.
\bibitem{Mantero}G. Montani, C. Mantero, F. Bombacigno, F. Cianfrani and G. Barca, \Journal{\PRD}{99}{063534}{2019}.
\bibitem{Maggiore2021}M. Fadel and M. Maggiore, \Journal{preprint}{\hspace{-0.14cm}}{}{2021}.
\end{thebibliography}
\end{document}